\documentclass[a4paper,12pt]{article}
\usepackage{jheppub}
\usepackage{fontawesome5}
\newcommand{\gs}{g_\star}
\newcommand{\gss}{g_{\star s}}
\newcommand{\Trh}{T_\text{rh}}
\newcommand{\Tp}{T_\text{peak}}
\newcommand{\arh}{a_\text{rh}}
\def\sfrac#1#2{{\textstyle{#1\over #2}}}
\newcommand{\be}{\begin{equation}}
\newcommand{\ee}{\end{equation}}

\title{Sterile Neutrino Dark Matter as a Probe of Inflationary Reheating}
\author[]{James M. Cline}
\author[]{and Yong Xu}
\affiliation{McGill University Department of Physics \& Trottier Space Institute\\
3600 Rue University, Montréal, QC, H3A 2T8, Canada}
\emailAdd{jcline@physics.mcgill.ca}
\emailAdd{yong.xu6@mcgill.ca}

\abstract{Sterile neutrinos offer a minimal and testable explanation for dark matter (DM), with their radiative decay actively searched for in X-ray observations. We show that cold sterile neutrino DM can be efficiently produced  during  reheating from inflaton decays with a small branching ratio, ${\rm BR}\lesssim 10^{-4}$. This production mechanism opens regions of parameter space where the active-sterile mixing is small enough to evade current X-ray constraints while reproducing the observed DM abundance. We systematically map the viable parameter space in terms of the sterile neutrino mass, mixing angle, inflaton mass, reheating temperature, and branching ratio. We further demonstrate that sterile neutrino DM can serve as a probe of inflationary reheating, with future X-ray observations capable of yielding information on the inflaton mass and the reheating-temperature ratio $m_\phi/T_{\rm rh}$. For a given inflationary model in which the inflaton mass is known, this leads to a lower bound on the reheating temperature that is several orders of magnitude stronger than the existing bound from Big Bang Nucleosynthesis.
}
\begin{document}
% \begin{flushright}
% December 2025
% \end{flushright}
\maketitle

%%%%%%%%%%%%%%%%%%%%%%%%%%%%%%%%%%%%%
\section{Introduction and Motivation}
%%%%%%%%%%%%%%%%%%%%%%%%%%%%%%%%%%%%%

Sterile neutrinos, $\nu_s$, are among the most widely studied dark matter (DM) candidates~\cite{Dodelson:1993je,Shi:1998km,Abazajian:2001nj,Asaka:2005an}; for reviews see, e.g.,~\cite{Boyarsky:2009ix,Boyarsky:2018tvu,Dasgupta:2021ies}. 
In the early Universe, they can be produced via oscillations from Standard Model (SM) neutrinos, known as the Dodelson–Widrow mechanism (hereafter BDKDW)~\cite{Barbieri:1989ti,Kainulainen:1990ds,Dodelson:1993je}\footnote{Although DW were the first to apply it for dark matter production, the mechanism was elucidated in the earlier references}.  This production peaks at a characteristic temperature $T_{\rm peak}\simeq130\,(m_s/{\rm keV})^{1/3}\,{\rm MeV}$, where $m_s$ is the sterile neutrino mass. Resonantly enhanced production is also possible, either in the presence~\cite{Shi:1998km,Abazajian:2001nj} or absence~\cite{Alonso-Alvarez:2022uxp} of a primordial lepton asymmetry.

Sterile neutrinos are unstable and can decay radiatively through their mixing with SM neutrinos, producing an X-ray photon with energy $E_\gamma\simeq m_s/2$. X-ray observations thereby provide stringent constraints on sterile neutrino DM. Current searches with XMM-Newton~\cite{Borriello:2011un}, NuSTAR~\cite{Krivonos:2024yvm}, Chandra~\cite{Hofmann:2016urz}, INTEGRAL/SPI~\cite{Calore:2022pks}, and XRISM~\cite{Yin:2025xad} {practically exclude} the parameter space in which the BDKDW mechanism accounts for the full DM abundance, with future missions such as eXTP~\cite{Malyshev:2020hcc} and eROSITA~\cite{eROSITA:2012lfj} expected to further tighten these bounds. This  has motivated a broad class of extensions featuring secret neutrino interactions~\cite{Cline:1991zb, Archidiacono:2014nda, Bezrukov:2017ike, DeGouvea:2019wpf, Farzan:2019yvo, Cline:2019seo, Alonso-Alvarez:2021pgy, Bringmann:2022aim, Dev:2025sah}. 
Such interactions allow  smaller mixing, thereby evading X-ray constraints. Realizing these scenarios typically requires introducing new bosonic degrees of freedom beyond the SM.

In this work, we eschew neutrino self-interactions beyond those present in the SM, while including sterile-neutrino production via BDKDW oscillations during radiation domination.  In an inflationary framework, however, the thermal history includes a reheating phase, when the inflaton decays and populates the thermal bath. This can provide
an additional source of sterile neutrino production directly from decays. 

A key parameter for determining the particle production is the reheating temperature, $T_{\rm rh}$, defined as the temperature at the onset of radiation domination following reheating. Currently, it is only constrained by Big Bang Nucleosynthesis (BBN), which implies $T_{\rm rh} \gtrsim \mathcal{O}({\rm MeV})$~\cite{Kawasaki:2000en,Hannestad:2004px,Barbieri:2025moq}. This bound is many orders of magnitude below typical inflationary scales and thus leaves the detailed dynamics of reheating essentially unconstrained.

The scarcity of messengers from the reheating era makes it  challenging to extract detailed information about its
properties.  
Photons and active neutrinos were tightly coupled to the thermal plasma throughout this epoch, preventing them from serving as effective probes. Future observations of gravitational waves produced during reheating offer a window 
\cite{Bernal:2023wus,Choi:2024ilx,Xu:2025wjq}, 
and  DM might also yield some information
about the reheating era.
For example, Ref.~\cite{Feldstein:2013uha} pointed out that direct detection of DM too heavy to be produced thermally would bound $T_{\rm rh}$ from below.   See Refs.~\cite{Bernal:2022wck,Becker:2023tvd,Gan:2023jbs,Xu:2023lxw,Boddy:2024vgt,Barman:2024nhr,Bernal:2024ndy,ShamsEsHaghi:2025kci} for other examples of DM inferences of reheating.

Motivated by these considerations and by the rapidly advancing  X-ray searches for sterile neutrino DM, we propose sterile neutrinos as a potential messenger of inflationary reheating. We quantify their production during and after reheating, including thermal contributions from oscillations and nonthermal ones from inflaton decays~\cite{Shaposhnikov:2006xi,Yaguna:2007wi,Bezrukov:2014nza,Chen:2025sgd}. If the inflaton is a gauge singlet, a Yukawa coupling $\phi\,\bar{\nu}_s\nu_s$ is allowed, leading to sterile neutrino production during reheating. Other production mechanisms have been studied in Refs.\ \cite{Kusenko:2006rh, Petraki:2007gq, Merle:2013wta, Adulpravitchai:2014xna, Merle:2015oja,Drewes:2015eoa, Shakya:2015xnx,DeRomeri:2020wng,Berbig:2022nre,Coy:2022unt,Lebedev:2023uzp, Koutroulis:2023fgp,Koivunen:2024vhr,Benso:2024qrg,Fuyuto:2024oii,Feiteira:2025phi} including freeze-in, freeze-out, and gravitational production.

In this paper, we analyze the interplay between oscillations and inflaton decays in determining the $\nu_s$ DM abundance, and we derive the viable regions in the space of $\nu_s$ mass and mixing angle, inflaton mass, reheating temperature, and inflaton branching ratio into sterile neutrinos. We demonstrate that $\nu_s$ DM yields information about the $m_\phi/T_{\rm rh}$ ratio, and that future X-ray observations have the potential to set lower bounds on $T_{\rm rh}$ that are several orders of magnitude stronger than BBN arguments.

The structure of this article is as follows. In Sec.~\ref{sec:Production}, we present the framework for computing sterile neutrino production, with emphasis on the phase-space distributions arising from oscillations and inflaton decays. Our main results are shown in Sec.~\ref{sec:Results}, where we also illustrate how sterile neutrino DM can be used to probe reheating. We summarize our findings in Sec.~\ref{sec:Conclusion}. In Appendix ~\ref{sec:App0}, the solutions for energy densities are presented. In Appendix~\ref{sec:appA}, we provide details on computing the collision term for sterile neutrino production from inflaton decays. We also present examples illustrating the evolution of neutrino phase-space distribution functions in Appendix~\ref{sec:appB}.

%%%%%%%%%%%%%%%%%%%%%%%%%%%%%%%%%%%%%
\section{Sterile Neutrino Production}\label{sec:Production}
%%%%%%%%%%%%%%%%%%%%%%%%%%%%%%%%%%%%%

In this section, we investigate the production of sterile neutrinos from the end of inflation until radiation domination after reheating. To this end, we first revisit the evolution of the cosmological background after inflation.
%%%%%%%%%%%%%%%%%%%%%%%%%%%%%%%%%%%%%
\subsection{Background Evolution}
%%%%%%%%%%%%%%%%%%%%%%%%%%%%%%%%%%%%%
After inflation ends, the inflaton field rolls toward the minimum of its potential and begins oscillating around it, transferring energy to other degrees of freedom. This process is known as reheating~\cite{Allahverdi:2010xz, Amin:2014eta}. In general, the inflaton can decay into Standard Model (SM) particles, such as the Higgs boson, as well as into particles beyond the SM. In this work, we do not specify the dominant decay channels; instead, we parametrize the total inflaton decay rate by $\Gamma_\phi$ and introduce a small branching ratio into sterile neutrinos, $\text{BR}(\phi \to \nu_s \nu_s)\ll 1$.

During reheating, the Universe consists of the inflaton field with energy density $\rho_\phi$, a radiation bath with energy density $\rho_R$, together with the active--sterile neutrino system. The radiation energy density is defined as
\begin{align}
    \rho_R(T) &= \frac{\pi^2}{30}\, \gs(T)\, T^4 ,
\end{align}
where $\gs(T)$ denotes the effective number of relativistic degrees of freedom contributing to the total radiation energy density. {Throughout this work, we assume that the inflaton oscillating around a quadratic potential $V(\phi) = \frac{1}{2}m_\phi^2 \phi^2$ during reheating, where $m_\phi$ corresponds to the inflaton mass.  Such a potential can arise in viable inflationary scenarios, including attractor inflation models \cite{Kallosh:2013hoa, Ellis:2013nxa}, Starobinsky inflation \cite{Starobinsky:1980te}, and polynomial inflation models \cite{Drees:2021wgd, Drees:2022aea}. The background evolution is governed by the Boltzmann equations
\begin{align}
  \dot{\rho}_\phi + 3H\rho_\phi &= - \Gamma_\phi \, \rho_\phi , \label{eq:rho_phi}\\
  \dot{\rho}_R   + 4H\rho_R   &= + \Gamma_\phi \,(1 - \text{BR})\, \rho_\phi ,  \label{eq:rho_R}
\end{align}
with the Hubble expansion rate given by
\begin{align}\label{eq:Hubble}
    H &= \sqrt{\frac{\rho_{\rm total}}{3 M_{P}^2}} \,,
\end{align}
where $\rho_{\rm total}$ denotes the total energy densities of the system, and $M_P$ denotes the reduced Planck mass.  Throughout this work, we assume $\text{BR} \ll 1$, so that the dominant fraction of the inflaton energy is transferred into radiation rather than into sterile neutrino DM. We define the end of reheating as the epoch at which the scale factor reaches $a = \arh$, where the inflaton and radiation energy densities satisfy
\begin{align}
\rho_\phi(\arh) = \rho_R(\arh) = \frac{3}{2} H^2(\arh) M_P^2
\equiv \rho_{\rm rh}\, .
\label{rhorheq}
\end{align}
This corresponds to a time scale $t(\arh) \simeq 1/\Gamma_\phi$, at which point most of the inflaton background has decayed. The temperature at this  point is defined to be the reheating temperature, $\Trh$.

We denote the scale factor at the beginning of reheating by $a = a_I$, while $a_{\rm rh}$ corresponds to that at the end. During the reheating phase, $a_I < a < \arh$, the total energy density is dominated by the inflaton,
$\rho_\phi \gg \rho_R$. As a consequence, the Hubble rate scales as $H \propto a^{-3/2}$, while the temperature evolves as $T \propto a^{-3/8}$, as follows from Eqs.~\eqref{eq:rho_phi} and \eqref{eq:rho_R}. More details are presented in appendix \ref{sec:App0}.
% This can be seen by introducing the comoving
% radiation energy density $E_R \equiv \rho_R a^4$; \textcolor{blue}{see more details in Appendix \ref{sec:App0}.}  It follows from Eq.~\eqref{eq:rho_R} that
% \[
% \frac{d E_R}{da} = \frac{\Gamma_\phi \rho_\phi a^3 }{H} \propto a^{3/2} \implies  E_R \propto a^{5/2} \implies \rho_R \propto a^{-3/2} \implies T \propto a^{-3/8}
% \]
% during reheating, assuming that $\rho_\phi \propto a^{-3}$ and $\Gamma_\phi$ is constant.\footnote{More generally, $\Gamma_\phi$ could depend on the amplitude of the inflaton oscillations, leading to different powers of $a$, depending on the details of the interactions; see for example Ref.\ \cite{Nurmi:2015ema}.  For simplicity we assume that $m_\phi$ and $\Gamma_\phi$ remain constant during reheating.}
It is important to notice that the maximum temperature attained during reheating, $T_{\text{max}}$, can be significantly larger than the nominal reheating temperature $\Trh$ \cite{Giudice:2000ex}.

With the background evolution specified above, we now turn to the study of the production and evolution of the sterile neutrinos.
%%%%%%%%%%%%%%%%%%%%%%%%%%%%%%%%%%%%%
\subsection{Boltzmann Equations for Neutrinos}
%%%%%%%%%%%%%%%%%%%%%%%%%%%%%%%%%%%%
In this section we present the framework for tracking the evolution of active-sterile neutrino system. Let $f_\alpha(p,t)$ denote the distribution function of active
neutrinos ($\alpha = e,\mu,\tau$) and $f_s(p,t)$ the sterile one.  Our aim to compute $f_s$, which is governed by the  semiclassical Boltzmann equation,
\begin{align}  \label{eq:fs}
  \left(\frac{\partial}{\partial t} - Hp\frac{\partial}{\partial p} \right)
  f_s(p,t)
  =
   \mathcal{C}_{s}\,,
\end{align}
where $H$ is given by Eq.~\eqref{eq:Hubble}, and $\mathcal{C}_{s}$ the collision term. For sterile neutrino production from oscillation, Eq.\ (\ref{eq:fs})  corresponds to the approximation of the full quantum kinetic equation in the small mixing regime \cite{Bell:1998ds,Johns:2019hjl}. With the phase space distribution function, one can further obtain the sterile neutrino number density
\begin{align}
  n_s (a) =\frac{g_s}{2\pi^2}\int_0^\infty dp\, p^2 f_s (p, a)\,,
\end{align}
where $g_s$ denotes the number of sterile neutrino degrees freedom; we will assume $g_s=2$ throughout this work, corresponding to a Majorana particle.
Finally, one obtains the abundance,
\begin{align} 
  Y_s = \frac{n_s}{s}\,,
\end{align}
where, $ s = \frac{2\pi^2}{45} \gss(T) T^3 $ denotes the entropy density, and $\gss$ counts the number of degrees of freedom contributing to the total entropy density. The present value must obey $Y_s m_s \simeq 4.3 \times 10^{-10}~\text{GeV}$ to match the observed DM abundance.

To determine the sterile neutrino distribution function $f_s$ and the corresponding yield $Y_s$, we consider two production mechanisms: $(i)$  production via active--sterile oscillations, and  $(ii)$ direct production from inflaton decays, as discussed in the following sections.

%%%%%%%%%%%%%%%%%%%%%%%%%%%%%%%%%%%%%
\subsubsection{Sterile Neutrinos from Oscillations}
%%%%%%%%%%%%%%%%%%%%%%%%%%%%%%%%%%%%%
The collision term in the Boltzmann equation, that describes
production of sterile neutrinos from active ones via oscillation,
is given by \cite{Abazajian:2001nj, Gelmini:2019wfp}
\begin{align}\label{eq:Cs_oscillation}
  \mathcal{C}_{\nu_\alpha \to  \nu_s}
  =
  \Gamma_\alpha
  \frac{1}{4}
  \sin^2 (2\theta_m)
 \left[
    f_\alpha (1-f_s) - f_s(1-f_\alpha)
  \right]\,.
\end{align}
Here, $f_\alpha$ ($f_s$)  is the phase-space distribution of active (sterile) neutrinos, while $\Gamma_\alpha$ is the active neutrino scattering rate, which we discuss below. Some treatments omit the $(1-f_s)$
Pauli-blocking  factor and the $-f_s(1-f_a)$ back scattering term  from $\mathcal{C}_{\nu_\alpha \to  \nu_s}$, but 
we will allow for the possibility that $f_s$ becomes sizable, in which case these effects must be taken into account. 

The parameter $\theta_m$  in Eq.~\eqref{eq:Cs_oscillation} is the in-medium mixing angle \cite{Abazajian:2001nj}:
\begin{align}\label{eq:mixing}
  \sin^2 (2\theta_m) =
  \frac{\Delta^2 \sin^2 (2\theta)}
       {(\Delta \cos2\theta - V_T)^2 + \Delta^2\sin^2 (2\theta) + D^2(p)}\,,
\end{align}
where $\Delta = \frac{m_s^2}{2p}$; the quantum damping rate is $D(p)= \Gamma_\alpha/2$.
{For definiteness, we will assume that $\nu_s$ mixes significantly only with the electron neutrino, justifying the
two-flavor approximation that is assumed in Eq.\ (\ref{eq:mixing}).
The thermal self-energy is given by \cite{Abazajian:2001nj}}
\begin{align}
  V_T \simeq -B\, p\, T^4\,,
\end{align}
where the coefficient $B$ has dimensions $\text{GeV}^{-4}$; we take its value as \cite{Abazajian:2001nj}
\begin{align}
B =
\begin{cases}
10.79\times10^{-9}\,\text{GeV}^{-4}\ (\text{e});\ 
3.02\times10^{-9}\,\text{GeV}^{-4}\ (\mu,\tau);\ T\lesssim 20\,\text{MeV}, \\[0.5em]
10.79\times10^{-9}\,\text{GeV}^{-4}\ (\text{e},\mu);\ 
3.02\times10^{-9}\,\text{GeV}^{-4}\ (\tau);\ 20\,\text{MeV}\lesssim T\lesssim 180\,\text{MeV}, \\[0.5em]
10.79\times10^{-9}\,\text{GeV}^{-4}\ (\text{e},\mu,\tau);\quad T\gtrsim 180\,\text{MeV}.
\end{cases}
\end{align}
Here the lepton flavors in parentheses denote which charged leptons are present in the plasma, contributing to the
$\nu_e$ thermal self-energy.
These formulae are valid at temperatures below the weak scale, where propagator effects of $W$ exchange are negligible.  The different behavior of $V_T$ at higher $T$ will have little effect on $\nu_s$ production, since the production peaks at temperatures well below the weak scale (see below). 

To study the evolution of $f_s$, we require the phase-space distribution function of active neutrinos, $f_\alpha$, which enters the collision term in Eq.~\eqref{eq:Cs_oscillation}. 
% We note that the background evolution starts at the end
% of inflation, when the temperature is still too low for $f_\alpha$ to be in
% equilibrium. For this reason, we do not assume that $f_\alpha$ is a simple Fermi-Dirac distribution, $f^{\text{eq}}_{\alpha} = \frac{1}{\text{e}^{p/T} +1}$.
We track the evolution of $f_\alpha$ through the Boltzmann equation
\begin{align}  \label{eq:fa}
  \left(\frac{\partial}{\partial t} - Hp\frac{\partial}{\partial p} \right)
  f_\alpha(p,t)
  =
   \mathcal{C}_{\alpha}
  -
   \mathcal{C}_{s},
\end{align}
Here, $\mathcal{C}_{\alpha} = \Gamma_\alpha(p,T)\bigl(f^{\text{eq}}_{\alpha} - f_\alpha\bigr)$ denotes the collision term describing active neutrino production from Standard Model scatterings, while $\mathcal{C}_{s}$ is given in Eq.~\eqref{eq:Cs_oscillation}. The equilibrium distribution $f^{\text{eq}}_{\alpha}$ is the Fermi--Dirac distribution, $f^{\text{eq}}_{\alpha} = \frac{1}{\mathrm{e}^{p/T} + 1}$. For temperatures $T \ll m_W$, the scattering rates reduce to the standard four-fermion approximation:
\begin{align}\label{eq:Gamma_alpha}
  \Gamma_\alpha(p,T)
  \simeq
  c_\alpha\, G_F^2\, T^4\, p \,,
\end{align}
where $c_e \simeq 1.13$, $c_{\mu,\tau} \simeq 0.79$ \cite{Gelmini:2019wfp}, and $G_F$ is the Fermi constant.  
Using the background evolution from Eqs.~\eqref{eq:rho_phi} and \eqref{eq:rho_R}, we can then study the evolution of the sterile--active neutrino system due to oscillations. In general, Eqs.~\eqref{eq:fs} and \eqref{eq:fa} cannot be solved analytically. Since $\mathcal{C}_{\alpha} \gg \mathcal{C}_{s}$, active neutrinos rapidly approach the thermal distribution due to the large scattering rate in Eq.~\eqref{eq:Gamma_alpha}. However, the thermal potential suppresses their conversion to sterile neutrinos at high temperatures, causing the sterile neutrinos to typically follow a nonthermal distribution. The production peaks when $|V_T| \sim \Delta$, corresponding to a temperature $T_{\text{peak}} \simeq 130 \left(\frac{m_s}{\rm keV}\right)^{1/3} \, \text{MeV}$ \cite{Kainulainen:1990ds}.
% Maybe show some examples for evolution of phase space? 
% Show some analytically approxmiation?
%%%%%%%%%%%%%%%%%%%%%%%%%%%%%%%%%%%%%
\subsubsection{Sterile Neutrinos from Inflaton Decay}\label{sec:decay}
%%%%%%%%%%%%%%%%%%%%%%%%%%%%%%%%%%%%%
When the inflaton decays $\phi\to\nu_s\nu_s$ are considered, there is an additional channel for producing sterile neutrinos.
The collision term for $\phi \to \nu_s \nu_s$  is given by\footnote{Further details are given in appendix \ref{sec:appA}.}
\begin{align}\label{eq:Cs_decay}
\mathcal{C}_{\phi \to \nu_s \nu_s}(p) = \frac{8 \pi^2 \,  n_\phi \Gamma_\phi \text{BR}} {m_\phi^2} \delta\left( p -\frac{m_\phi}{2}\right)\,.
\end{align}
Recall that $\text{BR}$ denotes the branching fraction into sterile neutrinos, and $n_\phi \equiv \rho_\phi/m_\phi$ is the inflaton number density. Since this is a two-body decay, all sterile neutrinos are produced with momentum $p = m_\phi/2$, as indicated by the $\delta$ function in Eq.~\eqref{eq:Cs_decay}. This feature makes the resulting distribution straightforward to solve analytically. 

The delta function in Eq.~\eqref{eq:Cs_decay} represents the instantaneous injection spectrum from inflaton decay in the homogeneous oscillating-background approximation. Possible broadening from inflaton momentum dispersion, rescattering, or preheating can be studied in specific reheating models, but these effects depend on the detailed microscopic dynamics of reheating and are not included in our model-independent treatment. The final sterile neutrino spectrum is nevertheless broadened by continuous production during reheating and by the redshift of particles produced at different times; this effect is included in Eq.~\eqref{eq:fs_decay_sol}  and in the numerical solution, as will be described below.

For convenience, we introduce the comoving momentum $q \equiv a\, p$, with which the Boltzmann equation can be rewritten as
\begin{align}
\frac{d f(q, a)}{d \ln a } = \frac{\mathcal{C}(q)}{H} \,.
\end{align}
For sterile neutrino production from inflaton decays alone, the phase-space distribution is given by
\begin{align}\label{eq:fs_decay_sol}
f_s(q, a) &= \int_{a_I}^{a} \frac{\mathcal{C}_{\phi \to \nu_s \nu_s}(q)}{a' H(a')} da' 
= \frac{8 \pi^2 \, \Gamma_\phi \, \text{BR}}{m_\phi^2} \int_{a_I}^{a} \frac{n_\phi(a')}{a' H(a')} \, \delta\left(\frac{q}{a'} - \frac{m_\phi}{2}\right) da' \nonumber \\
& \simeq 6 \sqrt{2} \, \pi^2 M_P^2 H_{\text{rh}} \Gamma_\phi\,\text{BR}  \sqrt{\frac{\arh^3}{m_\phi^5 q^3}} \quad \text{with} \quad a_I \leq \frac{2q}{m_\phi} \leq a \,,
\end{align}
where we have used $H \simeq H_{\text{rh}} (a/\arh)^{-3/2}$ and $n_\phi = \rho_\phi/m_\phi \simeq \rho_{\rm rh}/m_\phi \, (\arh/a)^3$ during reheating,\footnote{The scalings of $H$ and $n_\phi$ with $a$ are approximate, in that they ignore the effect  of the decay term
% the source term  going as 
% $e^{-\Gamma_\phi t}$
in the Boltzmann equation (\ref{eq:rho_phi}). Keeping that term leads to $\rho_\phi \propto  a^{-3} e^{-\Gamma_\phi t}$ as shown in Eq.~\eqref{eq:rhophi_sol}. The exponential suppression becomes significant at late time ($t\gtrsim 1/\Gamma_\phi$),  hence the following equations are approximations valid for $a\lesssim a_{\rm rh}$.
% that become accurate for $a\gg a_{\rm rh}$. 
} 
with $H_{\rm rh}$ denoting the Hubble parameter at $a = \arh$, and $\rho_{\rm rh}$ given by  using Eq.\ (\ref{rhorheq}). From Eq.~(\ref{eq:fs_decay_sol}), it follows that
\begin{align}
f_s(p, a) \simeq 6 \sqrt{2}\,  \pi^2 M_P^2 H_{\text{rh}} \Gamma_\phi \text{BR}\, \sqrt{\frac{\arh^3}{m_\phi^5 \, p^3 \, a^3}} \quad \text{with} \quad \frac{a_I}{a} \frac{m_\phi}{2} \leq p \leq \frac{m_\phi}{2} \,.
\end{align}
At the end of reheating, this reduces to
\begin{align}\label{eq:fs_arh}
f_s(\arh) \simeq 6 \sqrt{2}\, \pi^2 M_P^2 H_{\text{rh}} \Gamma_\phi \,\text{BR} \sqrt{\frac{1}{m_\phi^5 \, p_{\rm rh}^3}} \,,
\end{align}
where $p_{\rm rh}$ denotes the sterile neutrino momentum at the end of reheating, $p_{\rm rh} \equiv p_s(\arh)$; it lies in the range $\frac{a_I}{\arh} \frac{m_\phi}{2} \leq p_{\rm rh} \leq \frac{m_\phi}{2}$. With the phase-space distribution in hand, we are now ready to compute the number density.

At the end of reheating, the sterile neutrino number density is
\begin{align}\label{eq:ns}
n_s(\arh) &= \frac{g_s}{2 \pi^2} \int_{\frac{a_I}{\arh} \frac{m_\phi}{2}}^{\frac{m_\phi}{2}}dp_{\rm rh} \, p_{\rm rh}^2 \, f_s(\arh) 
\simeq \frac{2 M_P^2 H_{\text{rh}} \Gamma_\phi \, \text{BR}}{m_\phi} \nonumber \\
&\simeq 2 \, n_\phi(\arh) \, \text{BR} \,,
\end{align}
where we have taken $g_s = 2$ and used $\Gamma_\phi \simeq \frac{3}{2} H_{\rm rh}$, which follows from $H \simeq 2/(3t)$ during reheating and the definition of the reheating time scale $t \sim 1/\Gamma_\phi$. The final yield at present, $a = a_0$, is then 
\begin{align}\label{eq:Ys_decay}
Y_s(a_0) \simeq Y_s(\arh) \simeq \frac{3}{2} \left[\frac{\gs(\Trh)}{\gss(\Trh)} \right]\left(\frac{\Trh}{m_\phi}\right) \, \text{BR} \,,
\end{align}
assumimg entropy conservation after reheating. %\yx{Updates: In the previous email I wrote  $Y_s(a_0) \simeq \frac{\gss (a_0)}{\gss (\arh)} Y_s(\arh)$, which is wrong! $Y_s(a_0) \simeq Y_s(\arh)$ in Eq.~\eqref{eq:Ys_decay} is correct if we assume entropy conservation after reheating with $s(a_0) a_0^3= s(\arh)a^3_{\text{rh}}$.  Here is some more details (to be removed):
% \[
% Y_s(a_0) = \frac{ n_s (a_0)}{s(a_0)} = \frac{ n_s (\arh) \left(\frac{\arh}{a_0}\right)^3} {s(a_0)} =\frac{ n_s (\arh)} {s(a_0) a_0^3/a^3_{\text{rh}}} \simeq \frac{ n_s (\arh)} {s(\arh) } = Y_s(\arh).
% \]
% }  
With inflaton decays alone contributing to the $\nu_s$ production, it follows that 
 \begin{align}\label{eq:ms}
 m_s \simeq 28.7~\text{keV} \left(\frac{10^{-5}}{\text{BR}}\right)\left[\frac{\gss(\Trh)}{\gs(\Trh)} \right] \left(\frac{m_\phi}{\Trh}\right)\,,
 \end{align}
 by using $Y_s(a_0) m_s =4.3\times 10^{-10}~\text{GeV}$.
We have neglected the temperature dependence of the relativistic degrees of freedom in this analytical estimate, but in the following numerical analysis, their evolution is included. Moreover, in the analytical treatment we considered $\phi\to\nu_s\nu_s$ only from the end of inflation until the end of reheating, when $a = \arh$, whereas an additional contribution  may arise near $a=\arh$ after reheating, before the inflaton has completely decayed. This tends to reduce the numerical prefactor relative to that in Eq.~\eqref{eq:ms}. These effects are taken into account in Section \ref{sec:Results}.

%%%%%%%%%%%%%%%%%%%%%%%%%%%
\begin{figure}[!ht]
\def\sepf{0.8}
\centering
\includegraphics[scale=\sepf]{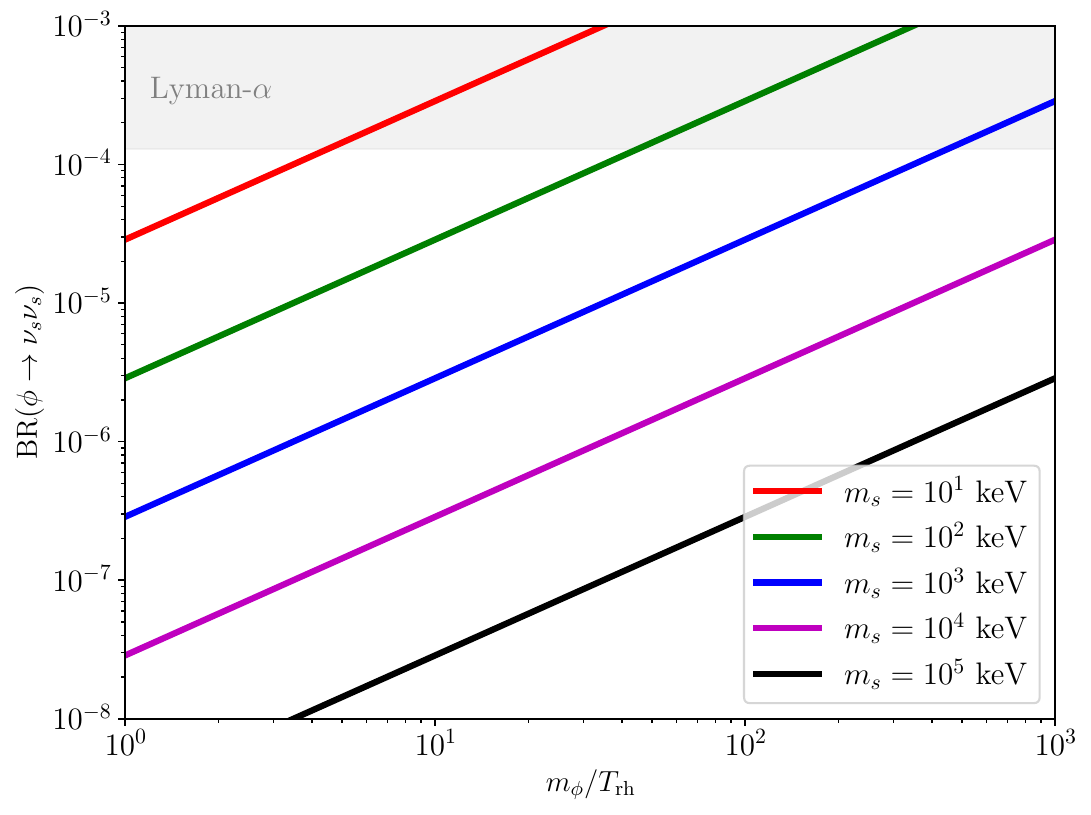}
\caption{Branching ratio for inflaton decay to sterile neutrinos (BR) as a function of $m_\phi/\Trh$, required to reproduce the observed DM abundance.}
\label{fig:BR}
\end{figure} 
%%%%%%%%%%%%%%%%%%%%%%%%%%%
To check whether the sterile neutrino dark matter is sufficiently cold, we compute its velocity at the present time,
\begin{align}\label{eq:vs0}
v_s (a_0) &= \frac{p_s(a_0)}{m_s} \simeq \frac{m_\phi}{2 m_s} \frac{\arh}{a_0} \simeq \frac{m_\phi}{2 m_s} \frac{T_0}{\Trh} \left[\frac{\gss(T_{0})}{\gss(\Trh)} \right]^{1/3}\,,
\end{align}
where $T_0 \simeq 2.73~\text{K}$ is the temperature of the CMB. 
We require $v_s$ to be smaller than current limits on the speed of warm DM. This implies $v_s (a_0)  <   1.8 \times 10^{-8}$,  based on Ref.\ \cite{Masina:2020xhk, Irsic:2017ixq} using Lyman-$\alpha$ constraints.  Combining Eqs.~(\ref{eq:ms},\,\ref{eq:vs0}) and  the upper bound on $v_s (a_0)$, we find 
% \begin{align}
% \frac{1}{2} \frac{T_0}{28.7~\text{keV}} \left[\frac{\gss(T_{0})}{\gss(\Trh)} \right]^{1/3}  \left(\frac{\text{BR}}{10^{-5}}\right)\left[\frac{\gs(\Trh)}{\gss(\Trh)} \right] < 1.8 \times 10^{-8}
% \end{align}
\begin{align}\label{eq:BR_bound}
\text{BR} \lesssim 1.3 \times 10^{-4}\,.
\end{align}
Ref.\ \cite{Bernal:2021qrl} found a similar result within a specific model of inflation. The estimate in Eq.~\eqref{eq:BR_bound} should be understood as an order-of-magnitude coldness criterion, rather than as a quantitative Lyman-$\alpha$ constraint. It is not used as an input to the abundance calculation or to the X-ray sensitivity analysis, but only as a guide for when the decay-produced sterile neutrinos are expected to be sufficiently cold.

In Fig.~\ref{fig:BR}, we show the branching ratio BR as a function of $m_\phi/\Trh$ which reproduces the observed dark matter abundance, assuming production solely from inflaton decay. The upper bound given in Eq.~\eqref{eq:BR_bound} is indicated by the gray band. We restrict ourselves to $\Trh \leq m_\phi$, as expected in typical perturbative reheating scenarios. The red, green, blue, magenta, and black curves correspond to sterile neutrino masses { $m_s = 10^{1},\,10^{2}\,\dots, 10^{5}~\text{keV}$, respectively.} As the sterile neutrino mass decreases, a larger branching ratio is required to account for the observed DM abundance.  {This simplified picture applies in the regime of small mixing angles, where production from 
$\nu_e$-$\nu_s$ oscillations can be ignored.  We next turn to the general case, where full numerical solution is required.}

To conclude this section, we clarify the scope of the analytic estimates derived above. We are assuming reheating around a quadratic minimum of the inflaton potential, for which the oscillating inflaton background admits a quasiparticle description with a well-defined mass $m_\phi$ and decay width $\Gamma_\phi$. This is the regime in which the mapping between $m_s$, $\text{BR}$, and $m_\phi/\Trh$ is direct and model-independent.

If the potential governing the post-inflationary oscillations is not quadratic, the estimates in Eqs.~\eqref{eq:fs_decay_sol}--\eqref{eq:BR_bound} generally receive corrections that depend on the reheating background. In such cases, the effective inflaton mass scale, decay rate, and sterile-neutrino injection spectrum are not fixed by the equation of state alone, but they also depend on the microscopic reheating channels. The analytic results presented above are therefore valid in the quadratic regime.

% %%%%%%%%%%%%%%%%%%%%%%%%%%%%%%%%%%%%%
% \subsubsection{Combined Analysis}
% %%%%%%%%%%%%%%%%%%%%%%%%%%%%%%%%%%%%%

%%%%%%%%%%%%%%%%%%%%%%%%%%%%%%%%%%%%%
\section{Combined Results}\label{sec:Results}
%%%%%%%%%%%%%%%%%%%%%%%%%%%%%%%%%%%%%

We now more quantitatively explore the parameter space $\{\theta,\,m_s\,,{\rm BR}\,,m_\phi\,,T_{\rm rh}\}$ that reproduces the observed dark matter (DM) abundance, taking into account both inflation decays and  $\nu_a\to \nu_s$ oscillations, by simultaneous numerical solution of Eqs.~\eqref{eq:rho_phi}, \eqref{eq:rho_R}, \eqref{eq:fs}, and \eqref{eq:fa}. The collision term $\mathcal{C}_s$ for $\nu_s$ production is given by the sum of $\mathcal{C}_{\nu_\alpha \to \nu_s}$ in Eq.~\eqref{eq:Cs_oscillation} and $\mathcal{C}_{\phi \to \nu_s \nu_s}$ in Eq.~\eqref{eq:Cs_decay}. The system is evolved from the end of inflation to $T=2\times10^{-2}\,\text{MeV}$ during the radiation-dominated era,\footnote{The results are insensitive to the exact final value taken for $T$.} 
 taking into account the $T$ dependences of $\gs$ and $\gss$. Examples of the evolution of the phase space distribution functions of $\nu_s$ are given in Appendix \ref{sec:appB}.  Numerical code used for this work is available on Github \href{https://github.com/yongxuDM/Sterile-Neutrino}{\faGithub}.\footnote{\url{https://github.com/yongxuDM/Sterile-Neutrino}}

%After we obtain the allowed parameter space, we also comment on how sterile neutrino experiments could help probe reheating.
%%%%%%%%%%%%%%%%%%%%%%%%%%%%%%%%%%%%%
\subsection{Parameter Space and Constraints}
%%%%%%%%%%%%%%%%%%%%%%%%%%%%%%%%%%%%%

%%%%%%%%%%%%%%%%%%%%%%%%%%%
\begin{figure}[!ht]
\def\sepf{0.8}
\centering
\includegraphics[scale=\sepf]{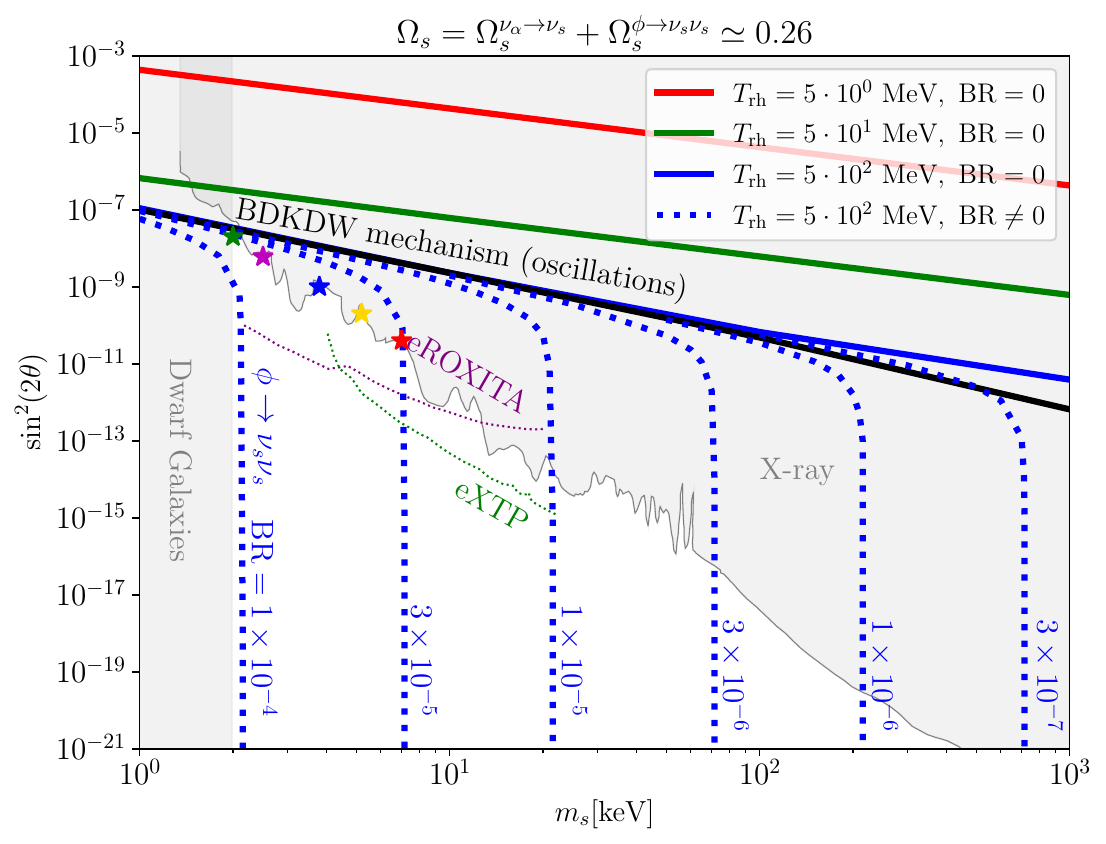}
\caption{Sterile neutrino DM parameter space $\{\sin^2(2 \theta), m_s\}$ for exemplary choices of BR and $\Trh$, assuming fixed inflaton mass
$m_\phi = 1$\,GeV.  Stars exemplify regions where BDKDW production is significant; see text for details.}
\label{fig:ps}
\end{figure} 
%%%%%%%%%%%%%%%%%%%%%%%%%%%

In Fig.~\ref{fig:ps}, we show mixing angles $\theta$ versus $\nu_s$ masses $m_s$
consistent with $\nu_s$ being all of the dark matter, $\Omega_s h^2 = 0.12$, and experimental constraints from X-ray observations, including XMM-Newton \cite{Borriello:2011un}, NuSTAR \cite{Krivonos:2024yvm}, Chandra \cite{Hofmann:2016urz}, INTEGRAL/SPI \cite{Calore:2022pks}, and XRISM \cite{Yin:2025xad}. The  projected sensitivities of forthcoming missions  eXTP  \cite{Malyshev:2020hcc} and eROSITA \cite{eROSITA:2012lfj}   are shown as narrow dotted curves, and constraints on the sterile neutrino dark matter mass from dwarf spheroidal galaxies \cite{Bezrukov:2025ttd} as the gray region to the left, assuming a fiducial value $m_\phi=1$\,GeV for the inflaton mass.
% Constraints from Lyman$-\alpha$ forest \cite{Garzilli:2019qki}. 
% update the plotes with  Lyman$-\alpha$ 

The red,  green, and blue lines correspond to scenarios with $\text{BR}=0$ and reheating temperatures  $5\times 10^{0}~\text{MeV}$,  $5\times 10^{1}~\text{MeV}$, and $5\times 10^{2}~\text{MeV}$, respectively\footnote{
The $\Trh=5\,{\rm MeV}$ curve is included as an illustrative oscillation-only comparison, showing the qualitative suppression of the BDKDW contribution when $\Trh<T_{\rm peak}$. It is not used in the inflaton-decay analysis or in deriving our main conclusions.
}. For high $T_{\rm rh}$, the predictions overlap with the black line, which corresponds to the standard BDKDW mechanism in a radiation-dominated Universe.  This behavior is expected, since even if $T_{\rm rh}$ is higher than the peak temperature $T_{\rm peak}$ of oscillation-induced production, the dominant contribution to $\nu_s$ still occurs at $T_{\text{peak}}$, assuming it happens after reheating.

However, if the reheat temperature is lower, $T_{\rm rh} < T_{\rm peak}$,
the final sterile neutrino abundance is determined by production at $\Trh$. In this case, although $\nu_s$ can be produced during reheating—when the temperature may temporarily  exceed $T_{\text{peak}}$—its abundance is subsequently diluted by entropy production. Consequently, the effective contribution remains controlled by $\Trh$, and a larger active–sterile mixing angle is required to compensate for the less efficient production, as illustrated by the green and blue curves. Fig.~\ref{fig:ps} underscores that the BDKDW mechanism cannot account for the observed dark matter abundance while remaining consistent with X-ray constraints.

The situation changes once inflaton decays into sterile neutrinos are included. As an example, we consider $m_\phi / \Trh = 2$ with $m_\phi = 1~\text{GeV}$.   The blue dotted curves determine the predicted values of $m_s$ as BR varies from  $3 \times 10^{-7}$   to $\text{BR} = 10^{-4}$, going right to left.  For example, with
$\text{BR} = 10^{-5}$, the sterile neutrino mass must be $m_s \simeq 21.5~\text{keV}$  to reproduce the observed DM abundance. We have included sterile neutrino production after reheating from inflaton decay, which increases the abundance relative to the approximation Eq.~\eqref{eq:Ys_decay}. Hence a smaller $m_s$ is needed compared to the value obtained in Eq.~\eqref{eq:ms}.

Since the inflaton decays are governed by the Yukawa coupling,
the active-sterile mixing can be arbitrarily small, thereby evading the X-ray bounds. As $m_s$ decreases along any of the dotted curves, the contribution to the total DM abundance from inflaton decay diminishes, requiring a larger contribution from the oscillations and causing the curve to bend. Eventually, it merges with the blue solid line, where most of the DM is produced by oscillations. %For smaller or larger branching ratios, a similar behavior is observed, as shown in 
% in Fig.~\ref{fig:ps}. 

{Although most of the allowed parameter space is dominated by inflationary production of $\nu_s$, there are some regions just below current X-ray bounds where a significant fraction of the $\nu_s$ dark matter is generated through
BDKDW oscillations.
Five such points are indicated by the green, magenta, blue, gold, and red stars, corresponding to
$(m_s/\mathrm{keV},\, \sin^2 2\theta)
= (2.0,\, 2\times10^{-8}),\;
(2.5,\, 6\times10^{-9}),\;
(3.8,\, 1\times10^{-9}),\;
(5.2,\, 2\times10^{-10}),\;
(7.0,\, 4\times10^{-11})$.
The associated contributions to the observed DM relic abundance are
$\Omega^{\nu_\alpha \to \nu_s}/\Omega_{CDM} \simeq
63.4\%,\; 27.9\%,\; 9.4\%,\; 5.5\%,\; 4.3\%$,
respectively. The contribution from the BDKDW mechanism is a decreasing function of $m_s$.}

In Fig.~\ref{fig:ps_Large_Trh}, we consider a larger inflaton mass,
$m_\phi = 10^{13}~\text{GeV}$, together with higher reheating temperatures
$\Trh$. The branching ratio is fixed to be $\text{BR}=10^{-4}$.
The blue dotted, dashed, and dash-dotted lines correspond to
$\Trh = 5 \times 10^{12}~\text{GeV}$,
$\Trh = 5 \times 10^{11}~\text{GeV}$, and
$\Trh = 5 \times 10^{10}~\text{GeV}$, respectively.
The qualitative behavior of these curves is the same as discussed above. In the regime where sterile neutrino DM production is dominated
by inflaton decays, the sterile neutrino mass scales as
$m_s \propto
\left(\frac{10^{-5}}{\text{BR}}\right)
\left(\frac{m_\phi}{\Trh}\right)$,
cf.~Eq.~\eqref{eq:ms}, demonstrating that only the ratio
$\left(m_\phi/\Trh\right)$ is relevant, and not the individual parameters.
As long as $\Trh > \Tp$, the sterile neutrino abundance produced
via oscillations is independent of $\Trh$, and the 
 results we show for illustrative
values of $m_\phi$ and $\Trh$ can easily be generalized to
other values.
%%%%%%%%%%%%%%%%%%%%%%%%%%%
\begin{figure}[!ht]
\def\sepf{0.8}
\centering
\includegraphics[scale=\sepf]{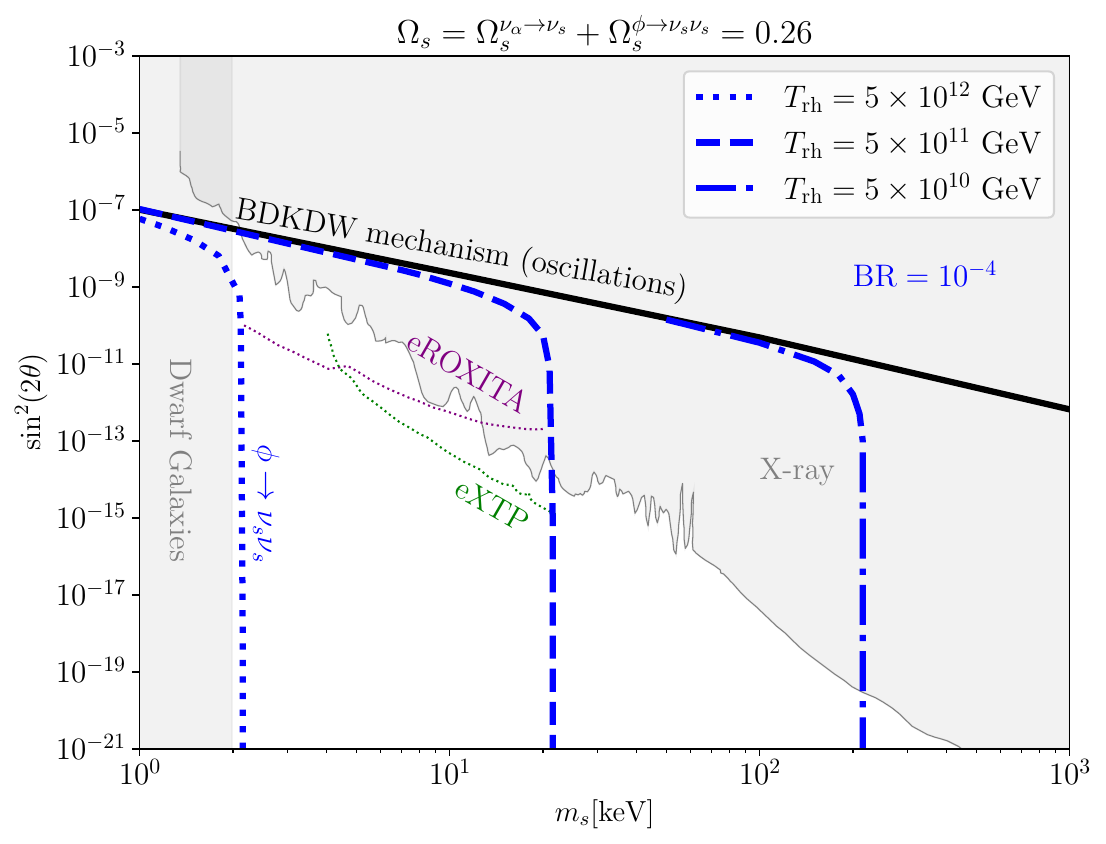}
\caption{Same as Fig.~\ref{fig:ps} but with larger $m_\phi$ and $\Trh$. }
\label{fig:ps_Large_Trh}
\end{figure} 
%%%%%%%%%%%%%%%%%%%%%%%%%%%

\subsection{Sterile Neutrinos as Probes of Reheating}
A novel implication of our work is that the discovery of 
 sterile neutrino DM could yield information about the details of
 reheating after inflation. This would be complementary to other scenarios that probe reheating through a dark sector, including scenarios involving WIMPs~\cite{Bernal:2022wck}, FIMPs~\cite{Becker:2023tvd,Barman:2024nhr}, axions~\cite{Xu:2023lxw}, millicharged particles~\cite{Gan:2023jbs}, dark photons~\cite{Bernal:2024ndy,ShamsEsHaghi:2025kci}, as well as  hypercharged DM ~\cite{Feldstein:2013uha}.

As shown in Figs.~\ref{fig:ps} and \ref{fig:ps_Large_Trh}, sterile neutrino production from inflaton decay can account for the entire DM relic abundance while remaining consistent with current X-ray constraints, with the allowed parameter space corresponding to the white regions in the figures. In this regime, the DM relics are determined by the inflaton branching ratio into sterile neutrinos and by the ratio $m_\phi/T_{\rm rh}$. Consequently, if sterile neutrinos are discovered---for instance through an X-ray signal consistent with $m_s \simeq 10~\mathrm{keV}$ and $\sin^2(2\theta) \simeq 10^{-13}$---such a measurement would map onto a narrow range of $m_\phi/T_{\rm rh}$ and $\mathrm{BR}(\phi \to \nu_s\nu_s)$, as shown in Fig.~\ref{fig:BR}. 
This should not be interpreted as an unambiguous reconstruction of the reheating history. Rather, within the reheating framework considered here, a sterile-neutrino signal would constrain the combination of $\mathrm{BR}(\phi\to\nu_s\nu_s)$ and $m_\phi/\Trh$ that controls the nonthermal abundance. If the inflaton mass is fixed by a specified inflationary model, or if $\mathrm{BR}$ is bounded or theoretically motivated, this constraint can be translated into information on $\Trh$. Imposing the bound $\text{BR} \lesssim 10^{-4}$, cf. Eq.~\eqref{eq:BR_bound}, one finds that $m_\phi/\Trh \lesssim 5$ for $m_s = 10~\text{keV}$.

As an example that is consistent with current inflationary constraints, suppose the Starobinsky model \cite{Starobinsky:1980te} describes the inflaton, and that it couples to the Higgs boson as $-\sfrac12\kappa\phi h^2$ for reheating.  Ref.\ \cite{Ellis:2025bzi} showed that radiative stability of the inflaton potential requires $\kappa < 4\times 10^{12}$\,GeV, corresponding to a reheat temperature $T_{\rm rh} < 4.2\times 10^{13}$\,GeV.  Further suppose that an X-ray line corresponding to $m_s=1$\,MeV sterile neutrino DM is observed; then Fig.\ \ref{fig:BR} constrains $m_\phi/T_{\rm rh} \lesssim 400$.  However, $m_\phi = 3\times 10^{13}$\,GeV is fixed by the COBE normalization in this model; therefore we can bound the reheat temperature as
\be
    7.5\times 10^{10}\,{\rm GeV} \lesssim T_{\rm rh} \lesssim 4.2 \times  10^{13}\,{\rm GeV}\,.
    \label{Trhest}
\ee
On the other hand, if reheating is principally into fermions, Ref.\ \cite{Ellis:2025bzi} obtains a bound of
$T_{\rm rh} < 2\times 10^{11}$ using cosmic microwave data from the Planck collaboration \cite{Planck:2018vyg}.  This considerably narrows the range in Eq.\ (\ref{Trhest}).

In a very different model, based on a renormalizable potential $V = b\phi^2 + c\phi^3 + d\phi^4$ with an inflection point, Ref.\ \cite{Drees:2021wgd} found upper limits $T_{\rm rh} <1\times 10^{11}$\,GeV ($4\times 10^8$\,GeV) for bosonic (fermionic) reheating, with $m_\phi\sim 1\times 10^{11}$\,GeV.   Taking the same $m_s=1$\,MeV as above, our results imply $2.5\times 10^8\,{\rm GeV} < T_{\rm rh}$.   These examples illustrate the potential for our scenario to provide lower bounds on the reheat temperature, that can be combined with upper bounds from technical naturalness to define an allowed window.

%%%%%%%%%%%%%%%%%%%%%%%%%%
\section{Summary and Conclusions}\label{sec:Conclusion}
%%%%%%%%%%%%%%%%%%%%%%%%%%
We performed a systematic study of sterile neutrino dark matter (DM) production, taking into account both thermal production via active–sterile neutrino oscillations—the Barbieri-Dolgov-Kainulainen-Dodelson-Widrow mechanism—and nonthermal production from inflaton decays during reheating. While the conventional scenario in which sterile neutrinos are produced solely via the BDKDW oscillations is strongly constrained by X-ray observations, we showed that a simple extension, in which the inflaton decays into sterile neutrinos with a small branching ratio, $\text{BR}\lesssim 10^{-4}$, opens up regions of parameter space that remain consistent with current X-ray limits.

Our main results are presented in Figs.~\ref{fig:ps} and~\ref{fig:ps_Large_Trh}. Allowing the inflaton to decay into sterile neutrino DM introduces an interplay between the two production mechanisms. In regimes with relatively large active–sterile mixing angles, the BDKDW mechanism dominates the production. We observed that when the reheating temperature exceeds the peak temperature of oscillation-induced production, the reheating dynamics do not affect the final DM abundance, thereby reverting to the standard BDKDW scenario.

In contrast, for sufficiently small mixing angles, sterile neutrino production is dominated by inflaton decay. In this regime, the DM mass is determined by the branching fraction of inflaton decay into sterile neutrinos, $\text{BR}$, and the ratio of the inflaton mass to the reheating temperature, $m_\phi/T_{\rm rh}$, as illustrated in Eq.~\eqref{eq:ms} and Fig.\ \ref{fig:BR}. Consequently, the mixing angle can be arbitrarily small, thus becoming compatible with bounds from X-ray observations. Therefore, a future positive detection of a sterile neutrino through its radiative decays would determine the ratio of the inflaton mass to the reheating temperature, thereby providing valuable information about the details of inflaton decay. For a given inflationary scenario in which the inflaton mass is known, this can in turn be translated into a lower bound on $T_{\rm rh}$.
%%%%%%%%%%%%%%%%%%%%%%%%%%
\section*{Acknowledgments}
%%%%%%%%%%%%%%%%%%%%%%%%%%%
This work was supported by the Natural Sciences and Engineering Research Council (NSERC) of Canada.
\appendix
\section{Solution for Energy Density}\label{sec:App0}
In this section,   we present analytical solutions for Eqs.~\eqref{eq:rho_phi} and \eqref{eq:rho_R}, which are used frequently in the main text. To factor out the Hubble expansion, we introduce the comoving energy densities $E_\phi \equiv \rho_\phi a^3$ and $E_R \equiv \rho_R a^4
$ \cite{Giudice:2000ex}. In terms of these variables, Eqs.~\eqref{eq:rho_phi} and \eqref{eq:rho_R} become

\begin{align}
\frac{d E_\phi}{da} &= -\frac{E_\phi}{a H} \Gamma_\phi \label{eq:Ephi}\, \\
\frac{d E_R}{da} &= +\frac{E_\phi}{H} \Gamma_\phi \label{eq:ER}\,.
\end{align}
From Eq.~\eqref{eq:Ephi}, we have
\begin{align} \label{eq:Ephi_sol}
  \ln \left[\frac{E_\phi (a)}{E_\phi (a_I)}\right] & \simeq \int^{a}_{a_I} -\frac{\Gamma_\phi}{a^{\prime}H_{I}(a^{\prime}/a_I)^{-3/2}} da^{\prime} \nonumber \\
  &= -\frac{2\Gamma_\phi}{3H_I} \left(\frac{a}{a_I}\right)^{3/2} \left[1 - \left(\frac{a_I}{a}\right)^{3/2}\right]\,,
\end{align}
where $H_I$ denotes the Hubble parameter at the beginning of reheating, and $H \simeq H_I (a/a_I)^{-3/2}$  during reheating. In general, $\Gamma_\phi$ could depend on the amplitude of the inflaton oscillations, leading to different powers of $a$, depending on the details of the interactions; see for example Ref.\ \cite{Nurmi:2015ema}.  For simplicity we assumed that $m_\phi$ and $\Gamma_\phi$ remain constant during reheating.
It follows from Eq.~\eqref{eq:Ephi_sol} that
\begin{align}\label{eq:rhophi_sol}
		\rho_\phi(a) &= \rho_\phi(a_I)\left( \frac{a_{I}}{a}\right)^{\! 3} \!\exp \left \{ -\frac{2\Gamma_\phi}{3H_I} \left(\frac{a}{a_I}\right)^{3\over 2}\!\! \left[1 - \left(\frac{a_I}{a}\right)^{3\over 2}\right] \right \}\nonumber \\
         &= \rho_\phi(a_I)\left( \frac{a_{I}}{a}\right)^3 \exp \left \{ -\frac{2\Gamma_\phi}{3H}  \left[1 - \left(\frac{a_I}{a}\right)^{3/2}\right] \right \}\, \nonumber \\
          & \simeq \rho_\phi(a_I)\left( \frac{a_{I}}{a}\right)^3 \exp \left \{- \Gamma_\phi t \left[1 - \left(\frac{a_I}{a}\right)^{3\over 2}\right] \right \}\,,
\end{align}
\noindent where $H \simeq \frac{2}{3t}$ was used in the last step. In regimes where $H \gtrsim \Gamma_\phi$ or equivalently $a \lesssim \arh $, the exponential term can be dropped. This implies during reheating, inflaton energy density scales as 
\begin{align}\label{eq:rhophi_sol2}
\rho_\phi(a) \simeq  \rho_\phi(a_I)\left( \frac{a_{I}}{a}\right)^3 \simeq \rho_\phi(\arh)\left( \frac{\arh}{a}\right)^3
\end{align}
in the regime $a_I< a \lesssim \arh$.
For $a >\arh$ or $t > 1/\Gamma_\phi$, the exponential suppression becomes significant, and consequently the inflaton energy density decreases substantially after reheating.

The solution to Eq.~\eqref{eq:ER} is
\begin{align}
E_R(a) - E_R(a_I)  & \simeq    \int^a_{a_I} \frac{E_\phi \Gamma_\phi}{ H_{I}(a^{\prime}/a_I)^{-3/2}} d a^{\prime} \nonumber \\
& =\frac{2}{5} \frac{E_\phi \Gamma_\phi}{H_I a_I^{3/2}} \left( a^{5/2}  - a_I^{5/2} \right)\,,
\end{align}
where $E_\phi$ remains constant, according to Eq.~\eqref{eq:rhophi_sol2}, and can therefore be taken out of the integral.
Assuming there is no radiation at the beginning of reheating, $E_R(a_I) =0$, we have 
\begin{align}\label{eq:rhoR_sol}
\rho_R (a) & \simeq \frac{2}{5} \frac{E_\phi \Gamma_\phi}{H_I a_I^{3/2} a^{3/2}} \left[  1 - \left(\frac{a_I}{a}\right)^{5/2} \right] \nonumber \\
& \simeq \frac{2}{5} \frac{\Gamma_\phi}{H_I } \rho_\phi(a_I)\left(\frac{a_I}{a}\right)^{3/2}\,.
\end{align}
In the last step, we took $a_I < a \lesssim a_{\rm rh}$ and neglected the term $\left(\frac{a_I}{a}\right)^{5/2}$. From the second line of Eq.~\eqref{eq:rhoR_sol}, it then follows that the temperature scales as $T \propto a^{-3/8}$ during reheating. Moreover, from the first line of Eq.~\eqref{eq:rhoR_sol}, $\rho_R(a)$ reaches its maximum at $a_{\rm max} = \left(\frac{8}{3}\right)^{2/5} a_I \simeq 1.5\, a_I.$
The temperature corresponding to $a_{\rm max}$ defines the maximum temperature $T_{\rm max}$, which can be significantly higher than the reheating temperature $T_{\rm rh}$ \cite{Giudice:2000ex}.

%
%%%%%%%%%%%%%%%%%%%%%%%%%%
 %\quad \text{for} \quad \frac{3H}{2} \gg \Gamma_\phi \quad \text{or}  \quad a \ll \arh \quad  \text{or}  \quad t \ll t(\arh) \nonumber

\section{Collision Term for Sterile Neutrino from Inflaton Decay}\label{sec:appA}

We start from the definition of number density for a particle specie $i$:
\begin{align}\label{eq:ni}
n_i \equiv \int d^3 \vec{p} \frac{g_i}{(2\pi)^3}f_i(\vec{p}) =4\pi\int dp\,p^2\, \frac{g_i}{(2\pi)^3}f_i(\vec{p}) \,,
\end{align}
with $g_i$ denoting degrees of freedom, $f_i$ representing the phase space distribution function, and $p \equiv |\vec{p}|$. The inflaton is taken to be a heavy particle at rest, $\vec{p}_\phi =0$. It follows from  Eq.~\eqref{eq:ni} that
\begin{align}\label{eq:fphi}
f(\vec{p}_\phi)
% = (2\pi)^3 \delta^3 (\vec{p}_\phi) 
\equiv  2\pi^2 \frac{n_\phi}{|\vec{p}_\phi|^2}\delta(|\vec{p}_\phi|)\,,
\end{align}
where  $g_\phi \equiv 1$. 

Knowing the phase space distribution for the inflaton, we can compute the collision term for the daughter particle, $\nu_s$.
We consider a process $\phi(p_\phi) \to \nu_s(p_1) \nu_s (p_2)$, where $p_\phi$, $p_1$, and $p_2$ denote the four momenta for the  inflaton, and the two sterile neutrinos. The collision term for producing  $\nu_s$  is given by 
\begin{align}\label{eq:Cs_f1}
\nonumber
\mathcal{C}_s[f_s(|p_1|)] &= \frac{1}{2E_1} \int d\Pi_\phi\,d\Pi_2 \,(2\pi)^4 \delta^4 (p_\phi - p_1 -p_2) 
\\ \nonumber 
&  \times 
\left\{f(\vec{p}_\phi)  \left[1-f_s(\vec{p}_2) \right] \left[1-f_s(\vec{p}_1) \right] -f_s(\vec{p}_1) f_s(\vec{p}_2) \left[1+f(\vec{p}_\phi) \right]\right\} |\mathcal{M}|^2  \nonumber \\
&\simeq \frac{1}{2E_1} \left(\int d\Pi_\phi f(\vec{p}_\phi) \right)\int d\Pi_2\, (2\pi)^4 \delta^4 (p_\phi - p_1 -p_2) |\mathcal{M}|^2  \nonumber \\
% &  =\frac{1}{2E_1} \frac{n_\phi}{2 \sqrt{m_\phi^2 +0^2}} \int \frac{d^3\vec{p}_2}{2E_2} (2\pi)\delta^3 (0 - \vec{p}_1 -\vec{p}_2)   \delta (m_\phi - E_1 -E_2)  |\mathcal{M}|^2  \nonumber \\
&  =\frac{1}{2E_1} \frac{n_\phi}{2 m_\phi} \int \frac{d^3\vec{p}_2}{2E_2} (2\pi)\delta^3 (0 - \vec{p}_1 -\vec{p}_2)   \delta (m_\phi - E_1 -E_2)  |\mathcal{M}|^2  \nonumber \\
&  =\frac{1}{2E_1} \frac{n_\phi}{2 m_\phi} \frac{(2\pi)}{2E_1}   \delta (m_\phi - 2E_1)   |\mathcal{M}|^2  %\nonumber \\ &
  =\frac{n_\phi \pi}{4E_1^2 m_\phi} \delta (m_\phi - 2E_1)  |\mathcal{M}|^2  \nonumber  \\ 
  & 
  \simeq \frac{8 n_\phi \pi^2  \Gamma_\phi \text{BR}}{m_\phi^2} \delta (|\vec{p}_1| -m_\phi/2) \,,
\end{align}
where $d\Pi_i = {d^3\vec{p}_i}/[{2E_i (2\pi)^3}]$ and Eq.~\eqref{eq:fphi} was used in the second step. In the second step,  Pauli blocking and back scattering terms are omitted since  $f_s < 1$, and $f_s < f_\phi$. In the last step $|\mathcal{M}|^2 \simeq y^2 m_\phi^2$ for the squared matrix element, where $y$ denotes the Yukawa coupling in $y \phi \bar{\nu}_s \nu_s$. The decay rate for the channel $\phi \to \nu_s \nu_s$ is then given by $\Gamma^{s}_\phi \simeq \frac{y^2 m_\phi}{16 \pi} = \text{BR} \, \Gamma_\phi$ with $\Gamma_\phi$ denoting the total inflaton decay rate, and $\text{BR}$ the branching ratio  into $\nu_s$. We  neglect $m_s$ since $m_\phi \gg m_s$; this leads to $E_1 = \sqrt{|\vec{p}_1|^2 + m^2_s} \simeq |\vec{p}_1|$ in the last step.

\section{Evolution of Neutrino Phase Space Distribution}\label{sec:appB}
%%%%%%%%%%%%%%%%%%%%%%%%%%%

%%%%%%%%%%%%%%%%%%%%%%%%%%%
\begin{figure*}[!ht]
\def\sepf{0.3}
\centering
\includegraphics[scale=\sepf]{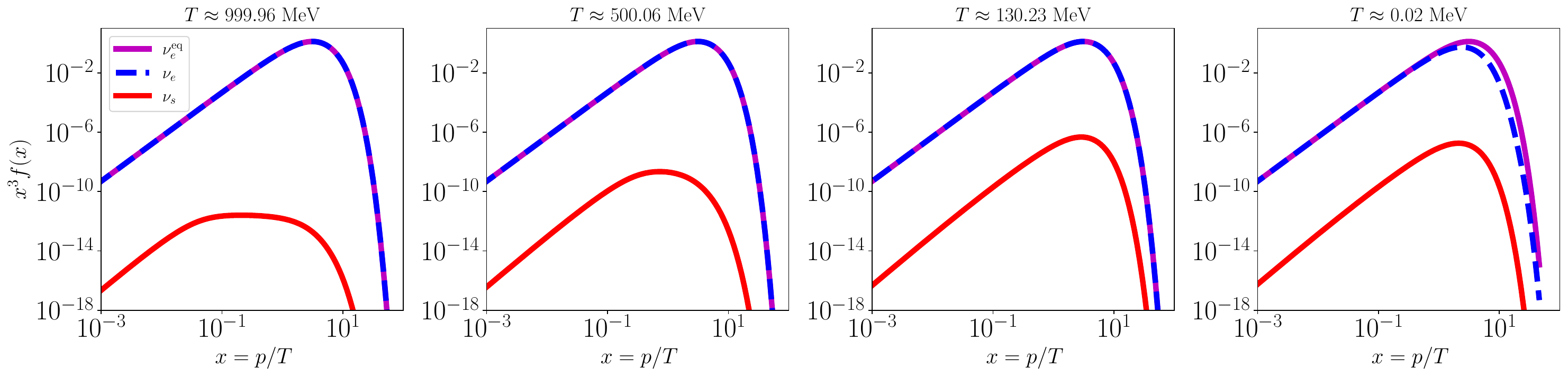}
\includegraphics[scale=\sepf]{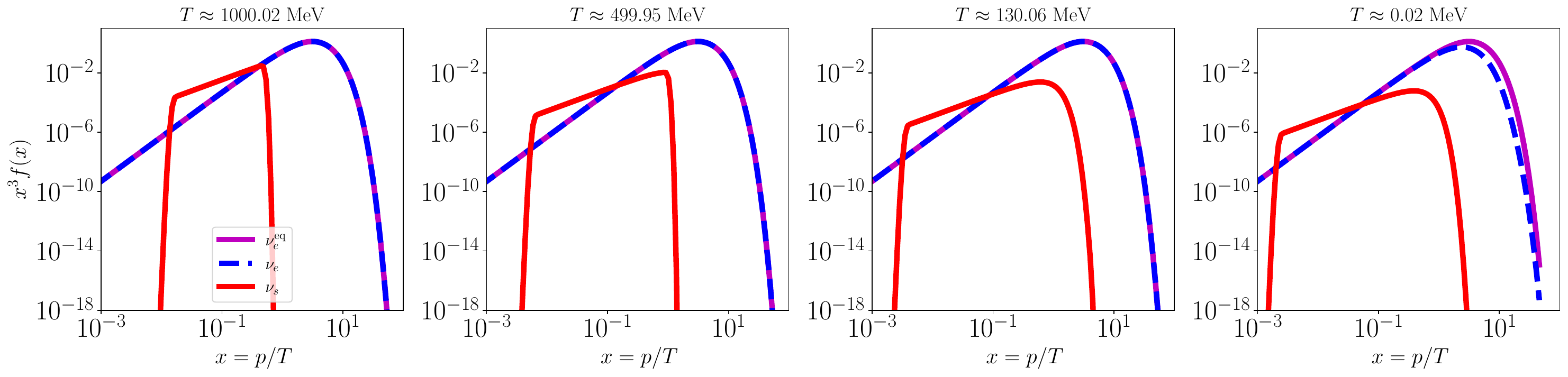}
\includegraphics[scale=\sepf]{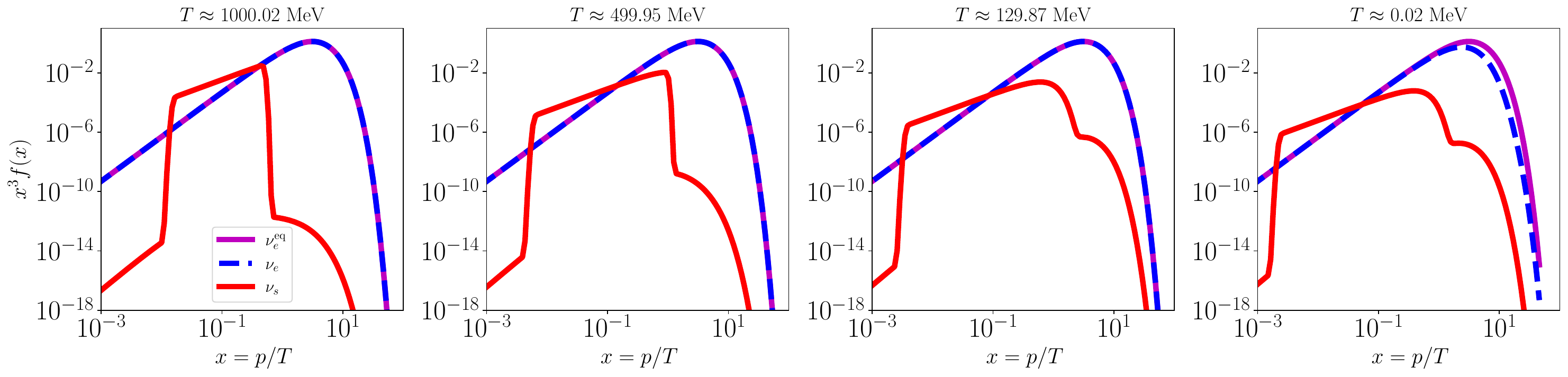}
\caption{Evolution of the neutrino spectra $x^{3} f(x)$ with decreasing temperature.
Rows correspond to different admixtures of oscillations and decays, as described in the text. }
\label{fig:f}
\end{figure*} 
%%%%%%%%%%%%%%%%%%%%%%%%%%%
In Fig.~\ref{fig:f}, we show the evolution of $x^{3} f(x)$ as a function of $x \equiv p/T$ for $m_s = 7~\text{keV}$, $m_\phi = 1~\text{GeV}$, $T_{\rm rh} = 500~\text{MeV}$, and $a_{\rm rh}/a_I \simeq 100$. The four columns correspond to temperatures $T \simeq 1000~\text{MeV}$, $T \simeq 500~\text{MeV}$, $T \simeq 130~\text{MeV}$, and $T \simeq 0.02~\text{MeV}$, respectively. The red solid, blue dashed, and magenta solid curves represent $\nu_s$, $\nu_e$, and the equilibrium distribution $\nu_e^{\rm eq}$, respectively. Below, we briefly explain the features of these plots. Numerical code used for this work is available on Github \href{https://github.com/yongxuDM/Sterile-Neutrino}{\faGithub}\footnote{\url{https://github.com/yongxuDM/Sterile-Neutrino}}.

\begin{itemize}
    \item \textbf{First row: pure oscillation} with
    $\sin^2(2\theta)=10^{-12}$ and $\mathrm{BR}=0$. At temperatures well above the MeV scale, active neutrinos remain in equilibrium, while sterile neutrino production is suppressed by matter effects in scattering processes. Consequently, $\nu_s$ typically stays out of equilibrium, as illustrated by the red curves. As the temperature drops, $f_s(x)$ initially increases and subsequently freezes in once the temperature falls below the peak production regime, $T \sim 130~\text{MeV}$. After that, it approaches a constant value at late times with small temperature. 

    \item \textbf{Second row: pure inflaton decay} with
    $\sin^2(2\theta)=0$ and $\mathrm{BR}=3\times10^{-5}$. For $\nu_s$ from inflaton decays, the initial momentum at production is $p \simeq m_\phi/2$, resulting in a sharp spectrum at early times during reheating. The spectrum broadens over time due to ongoing production and the expansion of the Universe during reheating. Finally, it becomes frozen once inflaton decays have completed. After reheating, the collision term becomes negligibly small,   since $n_\phi$  in Eq.~\eqref{eq:Cs_decay} is exponentially suppressed with $a>\arh$. This can also be seen from Eq.~\eqref{eq:rhophi_sol}.

    \item \textbf{Third row: combined effects} with
    $\sin^2(2\theta)=10^{-12}$ and $\mathrm{BR}=3\times10^{-5}$.
  This case illustrates the interplay between oscillation-induced production and inflaton decay. For the benchmark parameters considered here, the UV and IR parts of the spectrum are dominated by oscillations, while inflaton decay generates a pronounced peak in the intermediate momentum range. Both the location and height of this peak are determined by reheating parameters, such as $m_\phi$ and $T_{\rm rh}$, as also reflected in Eq.~\eqref{eq:fs_arh}. In this way, the phase-space distribution of $\nu_s$  encodes information about the reheating epoch.
\end{itemize}

\bibliographystyle{JHEP} 
\bibliography{biblio}
%%%%%%%%%%%%%%%%%%%%%%%%%
\end{document}